\documentclass[twocolumn,pra,preprintnumbers,amsmath,amssymb,superscriptaddress]{revtex4}

\usepackage{graphicx}

\begin{document}

	\title{Free-space spectro-temporal and spatio-temporal conversion for pulsed light}
	
	\author{E. Poem}
	\email{eilon.poem@weizmann.ac.il}
	\affiliation{Clarendon Laboratory, University of Oxford, Parks Road, Oxford OX1 3PU, United Kingdom} \affiliation{Department of Physics of Complex Systems, Weizmann Institute of Science, Rehovot 760001, Israel}
	\author{T. Hiemstra}
	\author{A. Eckstein}
	\affiliation{Clarendon Laboratory, University of Oxford, Parks Road, Oxford OX1 3PU, United Kingdom}
	\author{X.-M. Jin}
	\affiliation{Department of Physics, Shanghai Jiao Tong University, Shanghai 200240, PR China}
	\author{I. A. Walmsley}
	\affiliation{Clarendon Laboratory, University of Oxford, Parks Road, Oxford OX1 3PU, United Kingdom}
	
	\date{\today}

\begin{abstract}
We present an 
apparatus that converts every pulse of a pulsed light source to a pulse train in which the intensities of the different pulses are samples of the spatial or temporal frequency spectrum of the original pulse. In this way, the spectrum of the incident light can be measured by following the temporal response of a single detector. The apparatus is based on multiple round-trips inside a 2f-cavity-like mirror arrangement in which the spectrum is spread on the back focal plane, where after each round-trip a small section of the spectrum is allowed to escape. The apparatus is fibre-free, offers easy wavelength range tunability, and a prototype built achieves over 10\% average efficiency in the near infra red. We demonstrate the application of the prototype for the efficient measurement of the joint spectrum of a non-degenerate bi-photon source in which one of the photons is in the near infra red. 
\end{abstract}

\maketitle

\section{Introduction}\label{sect1}
Complex signals such as images and spectra are usually acquired by spatial multiplexing, using multiple detectors such as CCD or CMOS arrays. An alternative way of complex signal acquisition is temporal multiplexing, where signals from different parts of the image or spectrum arrive at a single detector at different times~\cite{Jalali_Review}. This allows for the use of detectors that are more sensitive and/or respond faster than CCD or CMOS arrays, such as fast photo diodes, avalanche photo diodes (APDs), or superconducting nano-wire photo detectors. Temporal multiplexing can therefore enable much faster acquisition of complex signals than traditional spatial multiplexing~\cite{Jalali_Review}. Furthermore, when the measurement of correlations between two or more complex signal channels is required, as, e.g. in Hanburry-Brown and Twiss interferometry~\cite{HBT_Sirius}, or in the spectral characterisation of a two-photon state~\cite{Silberhorn_JSI,Gerrits_HOM}, temporal multiplexing can be much more efficient than pixel-by-pixel correlation, as every event can be registered, and no events are discarded. 

Methods of conversion between spectral encoding and temporal encoding of information include dispersion in optical fibres~\cite{First_fibre_spect}, or in chirped fibre Bragg gratings (CFBGs)~\cite{First_cfbg,Jalali_800_cfbg}. These methods can also be used to convert between spatial and temporal encoding by first using standard dispersive optics such as a grating or a prism to convert between spatial and spectral encoding~\cite{Jalali_Review}. However, due to losses, the use of long fibres limits the spectral range to Telecom wavelengths only. CFBGs, while being much shorter, and therefore efficient also outside the Telecom range~\cite{Jalali_800_cfbg}, are limited to fixed, pre-determined wavelength ranges.

Here we present the SCISSORS (Spectro-temporal Conversion Interface by Successive Sectioning of Optical Radiation Sources) apparatus. SCISSORS converts information encoded in the spatial or spectral degree of freedom of pulsed light, to information encoded in the temporal degree of freedom. It is completely implemented it free-space, and uses no fibres. We show that in the near infra red (NIR), SCISSORS is at least as efficient as CFBGs, and with some modifications could significantly surpass them. It also offers additional features such as easy central-wavelength and range tunability, and can be adapted to operate efficiently in a wide range of wavelengths, making it appealing for various fast-acquisition microscopy applications~\cite{Jalali_Review}.

The manuscript is organized as follows: In Sect.~\ref{sect2} we describe the SCISSORS apparatus. In Sect.~\ref{sect3} we characterize a prototype of the apparatus and calibrate it. In Sect.~\ref{sect4} we use the prototype together with a fibre-based spectrometer for measuring the joint spectrum of a non-degenerate photon-pair source, utilizing the high efficiency of the prototype in the NIR. 
In Sect.~\ref{sect5} we discuss possible improvements of the prototype, as well as some additional possible applications for it, and in in Sect.~\ref{sect6} we conclude. 

\section{Apparatus design}\label{sect2}
\begin{figure}[tbh]
	\centering
\includegraphics[width=0.47\textwidth]{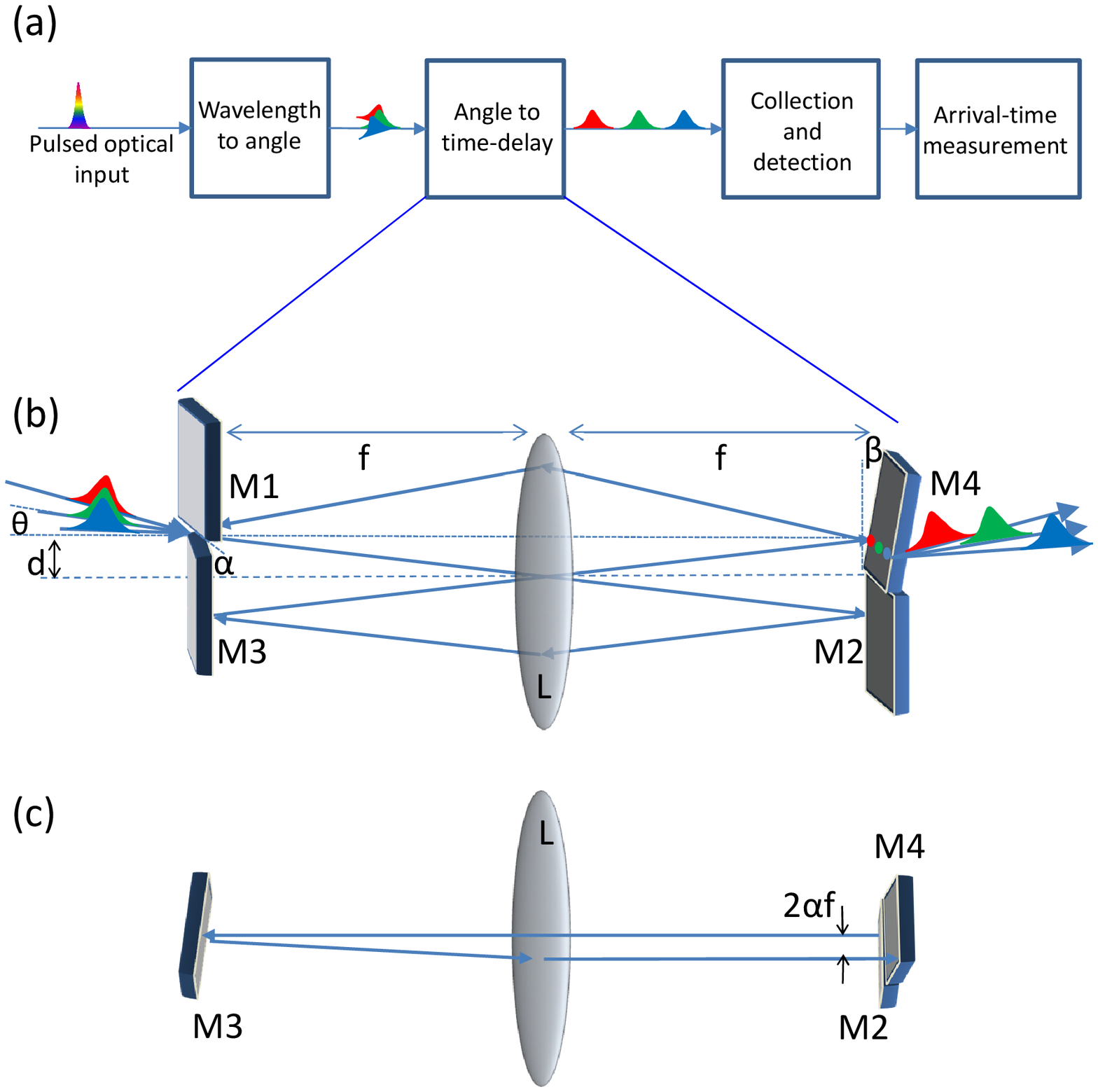}
 \caption{Schematics of SCISSORS. (a) Block diagram of the main components of the apparatus. (b) A detailed scheme of the angle to time converter (ATC). M1-M4, mirrors; L, a lens, with focal length $f$. The light enters between M1 and M3, at a vertical angle $\theta$, where all the colours are on the same spot, but in different horizontal angles. The vertical tilt of M4 prevents the light from escaping after one round trip through the entrance point, and enables multiple round trips. On M4, every colour is in a different position. (c) Top view of the ATC, mirror M1 removed for clarity. Due to the horizontal tilt of M3, every round trip a the horizontal position of a specific colour on M4 moves towards the edge by a distance of $2\alpha f$. Different parts of the spectrum therefore reach the edge of M4 and leaves the ATC after different number of round trips. In this way, a single, multicolour pulse at the input is divided into a train of pulses at the output, each of a different colour range, separated by one round trip time.}
 \label{fig1}
\end{figure}

SCISSORS is composed of three main parts, as schematically presented in Fig.~\ref{fig1}(a). The first part converts between wavelength and angle, by the use of a diffraction grating (a prism could be used as well), and a telescope which perfectly images the grating plane onto the entrance of the second part. The second part, which is the heart of the apparatus, is the angle to time-delay converter (ATC). Its details will be elaborated below. The third part is an imaging system collecting the light from the ATC onto a single detector. The detector signals are then time resolved either by standard analogue electronics (for a linear detector), or by digital time-tagging electronics (for a Geiger-mode detector). 

For converting angle into time, the ATC circulates the light inside a specially designed mirror arrangement. As shown in Fig.~\ref{fig1}(b), by the use of lens L, the light undergoes a spatial Fourier transform from the input plane (mirrors M1 and M3) to the output plane (mirrors M2 and M4), converting between angle and position. 

The light is input between M1 and M3, above the optical axis of the lens and at a vertical angle, such that after one pass through the lens it hits M2, after two it hits M3, and then it hits M4. As shown in the top view of Fig.~\ref{fig1}(c), due to the horizontal tilt of mirror M3, every time the light hits mirror M4, its horizontal position is shifted by $2\alpha f$, where $\alpha$ is the tilt angle of M3, and $f$ is the focal length of the lens. Because at the first time the light hits M4, every colour, entering the ATC at a different angle, is at a different position, every colour will take a different number of round-trips to reach the edge of M4 and escape the ATC. 

In this way, the initial multi-colour pulse is divided into a train of pulses, each containing a narrow colour range, set by the angle $\alpha$, and separated in time from the previous pulse by the round-trip time $\tau\gtrsim 8f/c$, where $c$ is the speed of light. The mirror M4 is vertically tilted by an angle $\beta$, to ensure that the light that comes back to the input plane hits mirror M1, and does not go back through the entrance after one cycle. Due to this vertical tilt, the output vertical angle increases by $2\beta$ every round trip. This may limit the number of pulses to hit the detector, due to the limited numerical aperture (NA) of the collecting optics. Nevertheless, as shown below, about twenty pulses could be collected in the present implementation. 

\section{Prototype characterization}\label{sect3}
For building a SCISSORS prototype we have used the following optical arrangements. 

For the first part, the output of a single mode fibre (NA=0.13) was collimated to a beam of about 2 mm in width by an 8~mm focal-length lens. The beam was then narrowed down to 0.35 mm in the vertical direction by a 6-fold demagnifying cylindrical telescope, and directed to a 1200 groves/mm diffraction grating at near-Littrow configuration, with diffraction efficiency of about 60\%. The grating plane was perfectly imaged on the entrance to the ATC by a regular telescope demagnifying by another factor of 6, creating a spot about 350~$\mu$m wide and 55~$\mu$m high.
   
The ATC was composed of one 60 mm focal length, bi-convex, anti-reflection coated, 2" lens (Thorlabs LB1723-B), and four highly reflecting mirrors (Layertec $\#$109879). The input and output mirrors (M1 and M4 in Fig.~\ref{fig1}(b), respectively) had one sharp edge, where in M1 the edge was horizontal, and in M4 vertical. 

The maximal spectral resolution, $\delta\lambda$, for this setup, for light centred about $\lambda_0$=800 nm, is $\delta\lambda$=$\lambda_0/N\approx0.33$ nm, where $N\approx$2400 is the number of grating grooves covered by the beam. Taking into account the 6-fold demagnifying telescope and the 60 mm focal length lens, the size of a diffraction-limited spot (containing the resolution limited wavelength range of 0.33 nm) at the output plane of the ATC is about 150 $\mu$m (horizontal) by 900 $\mu$m (vertical). The sharpness of the input and output mirror edges was found to be about 50~$\mu$m. In order to reduce relative scattering losses on the edge of mirror M4 to about 10\%, the tilt angle of mirror M3 was chosen such that the round-trip horizontal shift on mirror M4 was about 500~$\mu$m. This set the spectral bin size to about 1 nm.

The third part of the apparatus was composed of a 5-fold demagnifying telescope (input NA=0.08, output NA=0.39), imaging the output of the second part onto a 200 $\mu$m diameter multi-mode fibre (NA=0.39) coupled to a Perkin-Elmer silicon APD (180~$\mu$m in diameter). The use of a multimode fibre is necessary here since due to the vertical tilt of mirror M4, the output of the ATC consists of many optical modes, each at a slightly different angle [see Fig.~\ref{fig1}(b)].

The arrival times of photons to the APD were measured using digital time-tagging electronics (qu$\tau$ools quTAU), and analysed by a computer to form a histogram. 

\begin{figure}[tbh]
	\centering
	\includegraphics[width=0.47\textwidth]{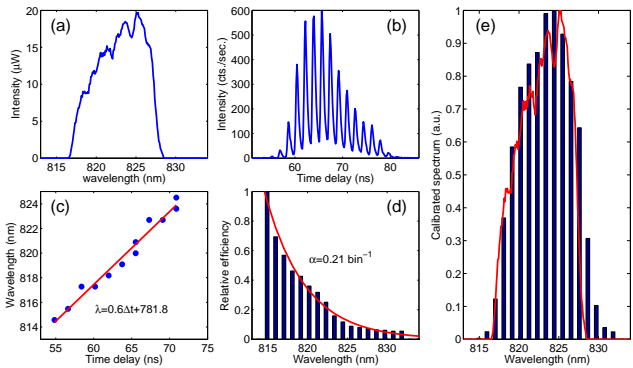}
	\caption{SCISSORS calibration. (a) An exemplary test spectrum, as measured on an OSA. (b) The output signal of the SCISSORS prototype, for the same input spectrum as in (a). (c) Time-delay to wavelength calibration curve, obtained by shifting the sharp spectral feature in the input spectrum and measuring the time where it appears at the output. (d) Relative efficiency calibration curve, obtained by comparing the added intensity between the OSA and the SCISSORS output for each shift of the spectral filter (bars). The line is an exponential fit, indicating 19\% loss per round-trip. (e) The SCISSORS output from (b), after wavelength calibration and efficiency correction (bars), compared to the input spectrum (line).}
	\label{fig2}
\end{figure}

For testing and calibrating the prototype, $\sim$50 fs pulses from a Ti:Sapph oscillator (Coherent Mira 900), amplified by a Ti:Sapph regenerative amplifier (Coherent RegA 9000) operating at 250 kHz repetition rate, where passed through a 10 nm band-pass spectral filter (Semrock FF01-830/2-25), to create sharp and tunable spectral features. A representative test spectrum, as recorded by an Ando AQ6317B optical spectrum analyser (OSA), is presented in Fig.~\ref{fig2}(a). Fig.~\ref{fig2}(b) shows the arrival-time histogram measured using the SCISSORS prototype for the input spectrum of Fig.~\ref{fig2}(a). About 15 distinct time bins are seen, separated from each other by 1.75 ns.  

By tilting the band-pass filter, the spectral feature was shifted in wavelength, which allowed for the calibration of the time-delay to wavelength ratio of the prototype. The results of this calibration are shown in Fig.~\ref{fig2}(c). The relative SCISSORS efficiency is calibrated by comparing the intensity added at every spectral shift between the OSA and the SCISSORS output. The results are presented in Fig.~\ref{fig2}(d) (bars). The relative efficiency fits rather well to a decaying exponential (line), indicating a loss of 19\% per spectral bin (i.e. per round trip). The spectrum obtained when the wavelength and efficiency calibrations where applied to the output SCISSORS signal is presented in Fig.~\ref{fig2}(e) (bars), and is compared to the input spectrum measured on the OSA (line). 

The average absolute efficiency of SCISSORS was measured for the spectrum shown in Fig.~\ref{fig2}(a) by comparing the total count rate measured at the SCISSORS output to that measured at its input, and was found to be $\sim$10$\%$. This sets the maximum absolute efficiency of the prototype to be $\sim$45$\%$. 

\section{Joint spectral intensity measurement of a non-degenerate bi-photon source}\label{sect4}
We demonstrated the operation of SCISSORS at very low light levels and at wavelengths not easily accessible to CFBG spectrometers by measuring the joint spectral intensity (JSI) of a pair of photons from a light source generating two-mode squeezed states. 
The source is a 4 cm long silica-on silicon UV-written waveguide, pumped at 1069 nm, emitting one photon around 1520 nm and the other around 825 nm. This source is based on birefringence phase-matched four-wave mixing (FWM)~\cite{Justin1}. For the pump beam, we used an optical parametric amplifier (Coherent OPA 9400) pumped by the Ti:Sapph regenerative amplifier. This system provided $\sim$70 fs pulses at 250 kHz repetition rate. Before coupling to the waveguide the spectrum of the pump was narrowed down to a width of 20 nm by a pulse shaper. The total average power coupled to the waveguide was 180 mW (pulse energy of 0.72 $\mu$J).

The photons emitted near 825 nm were coupled through a single-mode fibre to the SCISSORS, and eventually detected by a silicon APD. The photons emitted around 1520 nm were coupled to an $\sim$11 km long dispersion compensating fibre (DCF), where spectral information was converted to temporal information through the dispersion of the fibre, and eventually detected using an InGaAs APD (idQuantique id220).

\begin{figure}[tbh]
	\centering
	\includegraphics[width=0.48\textwidth]{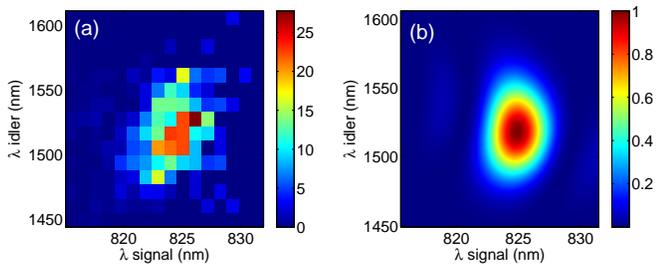}
	\caption{(a) The joint spectral intensity of a non-degenerate silica waveguide FWM photon pair source, as measured using a dispersive fibre (for the Telecom wavelengths) and the SCISSORS for the NIR. The colour bar shows the number of coincidence counts detected in 11 hours, after correcting by the measured SCISSORS relative efficiency curve [see Fig.~\ref{fig2}(d)], while keeping the total number of coincidences constant. (b) The predicted normalized JSI~\cite{Justin1}.}
	\label{fig3}
\end{figure}

The arrival times to both detectors with respect to the pump pulse were measured using the time-tagging electronics. If both detectors fired after the same pump pulse, the pair of arrival times was kept, and used for building up a two-dimensional histogram. This histogram, with the application of the calibrations of the two spectrometers, samples the JSI of the photon pair source. The measured JSI for the FWM source is presented in Fig.~\ref{fig3}(a), and the theoretical prediction~\cite{Justin1}, for birefringence of 3.36$\times$10$^{-4}$, is presented in Fig.~\ref{fig3}(b). The fidelity of the measured distribution (normalized), $P_m$, to the calculated one, $P_c$, given by \mbox{$F=\sum_i{P_m(x_i)P_c(x_i)}/\sqrt{\sum_i{P_m(x_i)^2}\sum_j{P_c(x_j)^2}}$}, where $x_{i(j)}$ is the $i^{th}$ ($j^{th}$) sampling point, is $F=91\%$. 

The measurement took 11 hours to complete. Taking into account the pair production rate of the source, we estimate that measuring its JSI, to the same spectral resolution and the same total number of coincidence events, using two traditional scanning gating spectrometers, each having 10\% total efficiency (including detector efficiency), would have taken about 300 hours.

\section{discussion}\label{sect5}
\subsection{Possible improvements}
The prototype presented here is capable of creating about 20 output time bins, with a differential efficiency ranging between 45\% and 5\%, and a spectral resolution of about 1 nm per bin.  This rapid decay of efficiency in each time bin is due to a rather high round-trip loss of 19\%. As the round-trip loss due to the nominal imperfect transmissivity of the lens and imperfect reflectivity of the mirrors is lower than 2\%, the main contribution to the round-trip loss must come from a different source. 

We suspect that near the edge of mirror M1, where most of the reflections off it occur, the reflectivity is reduced due to edge effects. These could be both scattering due to rough surface, or reduced reflectivity due to imperfect coating. The same effects also result in a one-off loss during the exit at mirror M4, which dictates a large horizontal shift and limits the spectral resolution of the prototype. We note that while the spectral resolution of the prototype is still lower than that achieved at the NIR using CFBGs~\cite{Jalali_800_cfbg}, its average efficiency ($\sim$10\%) is already comparable to that achieved with NIR CFBGs~\cite{Jalali_800_cfbg}.

One way to improve on the performance of the prototype would be by making the edges of mirrors M1 and M4 sharper. This may be achieved by using photo-lithography to make a mirror with a sharp, micron-scale transition between a high reflection area and a high transmission area. Apart from reducing the round-trip loss, such mirrors would allow the round-trip horizontal shift to be smaller, and thus increase the spectral resolution of SCISSORS for the same set of optics. They would also allow a lower vertical angle to be used, and more pulses to be contained within the NA of the collecting optics. This NA may also be increased by using a free-space coupled detector instead of the fibre-coupled one used here. We estimate that with these improvements alone, about 70 bins could be created with absolute efficiency between 5$\%$ and 50$\%$ and spectral resolution of less than 0.5 nm. 

The spectral resolution could be even further improved by changing the optics of the first stage. The absolute efficiency could be easily improved by using a more efficient grating, or a prism. In the later case, the first-stage optics could be adapted for compensating for the lower diffraction. These improvements would bring both the resolution and the efficiency of SCISSORS to be better than currently achieved at the NIR with CFBGs, whilest keeping its easy spectral range tunability.

 \subsection{Additional possible applications}
The heart of SCISSORS is the ATC, which converts angle to time-delay. One could therefore envision the use of the ATC as a part of a one-dimensional single pixel camera~\cite{Jalali_Review}. Another application for the ATC is the production of shaped pulse trains. This can be achieved by spatially shaping the input beam. Furthermore, since every pulse comes out of the ATC at a different angle, the produced pulse train could be used for illuminating different areas of an object at different times. Thus, combining two ATCs of different round-trip times, one for illumination, and one for collection, a two-dimensional single-pixel camera could be built, with no moving parts.
Additional applications for SCISSORS include hyper-spectral imaging (one dimensional with no moving parts, and two-dimensional by adding a scanning mirror), and temporal interferometry~\cite{Jalali_interf}, where it could replace the optical fibre as the time-stretch transformer. 
 
\section{Conclusion}\label{sect6}
We have built and tested a prototype of a novel apparatus for converting spectrally or spatially encoded optical information into temporally encoded information. The apparatus works by dividing an incoming pulse of light to a train of pulses, where every pulse in the output train contains light incident on the device at a specific section of angles. This is achieved by circulating the light in a specially designed mirror arrangement, where every section of input angles travels a different number of round-trips. The apparatus offers the possibility of high efficiency as well as large tunability of the spectral range converted and of the size of the spectral bins. We have used a prototype of this apparatus in combination with a fibre-based spectrometer to efficiently measure the joint spectrum of a non-degenerate bi-photon source. 
The apparatus could also be used for various classical applications, such as fast acquisition of images, the creation of shaped pulse trains, hyper-spectral imaging, and temporal interferometry.\\ 

This work was supported by the UK EPSRC (EP/J000051/1), the EU IP SIQS (600645), the US AFOSR EOARD (FA8655-09-1-3020), and an EU Marie Curie Fellowship (IEF-2013-627372 to EP). IAW acknowledges an ERC Advanced Grant (MOQUACINO) and an EPSRC Programme Grant (EP/K034480/1). EP acknowledges the British Technion Society for a Coleman-Cohen Fellowship, and the Oxford Martin School for initial support.\\

The authors would like to thank Dr. Brian Smith, Dr. Marco Barbieri, and Dr. Robert Evans for stimulating discussions.


\end{document}